\documentclass[%
 aip,
 jmp,%
 amsmath,amssymb,
 reprint,%
]{revtex4-1}
\usepackage{graphicx}
\usepackage{dcolumn}
\usepackage{bm}

\begin{document}

\preprint{AIP/123-QED}

\title[Relativistic Equation of State]{4-Velocity distribution function using Maxwell-Boltzmann's original approach and 
a new form of the relativistic equation of state.}

\author{Prasad Basu}
\affiliation{S. N. Bose National Centre For Basic Sciences, Salt-lake, Kolkata, INDIA.}
\email{prasadcsp@gmail.com}
\author{Soumen Mondal}%
\affiliation{Korea Astronomy and Space Science Institute, Daejeon, Republic of Korea 305-348.}
\affiliation{Ramakrishna Mission Residential College, Narendrapur, Kolkata, INDIA.}
\email{crabhorizon@gmail.com}

\date{\today}

\begin{abstract}
Following the original approach of Maxwell-Boltzmann(MB), 
we derive a 4-velocity distribution function for the relativistic ideal gas. 
This distribution function perfectly reduces to original MB distribution in the non-relativistic limit.
We express the relativistic equation of state(EOS), $\rho-\rho_0=(\gamma-1)^{-1}p$,\ in the two equations: 
$\rho=\rho_0 f(\lambda)$,\ and $p=\rho_0 g(\lambda)$, where $\lambda$\ is a parameter related to the kinetic energy, 
hence the temperature, of the gas.
In the both extreme limits, they give correct EOS:\ $\rho=3p$\ in the ultra-relativistic, and\ $\rho-\rho_0=\frac{3}{2}p$ 
in the non-relativistic regime. Using these equations the adiabatic index $\gamma$ (=$\frac{c_p}{c_v}$) and the sound speed 
$a_s$ are calculated as a function of $\lambda$. They also satisfy the inequalities: $\frac{4}{3} \le \gamma \le \frac{5}{3}$ 
and $a_s \le \frac{1}{\sqrt{3}}$ perfectly.
\end{abstract}

                              
\maketitle

\section{Introduction}
The thermodynamics of matter plays an important roll in the study of its fluid dynamics. The solution of hydrodynamic equations, both in the relativistic and non-relativistic formalism, require a knowledge about the equation of state (EOS) of the system.
An EOS is essentially a relation between the macroscopic quantities e.g. the pressure $p$, the total mass-energy density $\rho$, and rest mass energy density $\rho_0$. One could arrive into such a relation from the knowledge of the probability distribution of various microscopic quantities (e.g molecular velocity, energy). For the relativistic gas, such a study has been done using the formalism of statistical mechanics in a Lorentz covariant framework (See [\onlinecite{Synge1957}] and ref. therein, [\onlinecite{Chandrasekhar1939}], [\onlinecite{Tolman1934}], [\onlinecite{Taub1948}]). Distribution function of co-ordinate velocity ($\frac{dx}{dt}$) and the EOS can be derived using this formalism. Several authors tried to use relativistic EOS to solve the relativistic hydrodynamics (RHD) equation in various astrophysical problems ([\onlinecite{Shen1998}]). But a direct application ([\onlinecite{Falle1996}]) of the exact relativistic EOS turns out to be a less feasible method for numerical computing because of its complexity. Thus in the popular studies of numerical RHD in astrophysical problems (see review [\onlinecite{Wilson2003}]), people prefer to use various alternative models of EOS which can reproduce various features of exact EOS with a good approximation ([\onlinecite{Ryu2006}], [\onlinecite{Mignone2005}]).  All these models ([\onlinecite{Sokolov2001}]) are proposed in an ad-hoc manner and are not derived consistently from the first principle.

The present status of RHD study motivates us to analyze this problem in a model independent way in particular, we wish to find a simpler form of EOS which, apart from being computer friendly, must also be derived from very first principle of kinetic theory.  At this point, it would be interesting to see whether one could derive a probability distribution function of 4-velocity following the basic methods of probability theory and assumptions of isotropy as originally had been done by MB in the case of non-relativistic gas.This approach, to the best of our knowledge, is yet unexplored and one could examine it atleast for the sake of conceptual completeness. An EOS could be derived then using this distribution function.
In this article, we follow the original line of arguments of MB to find a distribution function of 4-velocity ($u^i=\frac{dx^i}{d\tau}$), instead of co-ordinate velocity $\frac{dx^i}{dt}$, for a relativistic ideal gas. We then use these Maxwell like 4-velocity distribution function to derive a form of EOS. 

  
This paper is organized as follows. In \S II, we derive Maxwell like 4-velocity distribution function for relativistic gas.
In \S III,  we apply it to find the EOS for relativistic gas. In \S IV, we investigate the behavior of some physical quantities (e.g. $\gamma$ and sound speed etc.). The extreme limits: ultra-relativistic and non-relativistic are discussed. Finally concluding remarks are drawn in \S V.

\section{Four velocity distribution of a relativistic ideal gas}
We consider a gas consisting of one species of non-interacting particles. The temperature of the gas is so high that the average kinetic energy of a constituent particle is comparable to its rest mass energy. In other words, the thermal energy of the gas is comparable with its rest mass energy.
The co-ordinate velocity and the 4-velocity of a gas particle are defined respectively as 
$v^{i}\!=\!\frac{dx^{i}}{dt}$ and $u^{\mu}\!=\!c\frac{dx^{\mu}}{d\tau}$  
where $d\tau^2\!=\!c^2dt^2\!-\!(d\vec x)^2$ is the length element in the Minkowski space-time.
\!$u^{\mu}$'s are related to the co-ordinate velocity as 
\begin{eqnarray}
u^{\mu}=c\frac{dx^{\mu}}{d\tau}=v^{\mu}/\sqrt{1-{{v^2}\over{c^2}}}, 
\label{umu}
\end{eqnarray}
where  $v^{\mu}=(1,\vec v)$.
The molecules of the gas interacts only through elastic collision 
so that all of their energy is kinetic energy. In a Lorentz frame, where the center of mass of the container 
is at rest, the distribution of molecular speed is isotropic.
The choice of such a frame is unique upto a three dimensional rotation. 
In different frames in which the container is at rest,  $u^0$ takes same value and $u^i$s are transformed  
by a 3-D Euclidean rotation {\it i.e} they transform like a components of an ordinary three dimensional 
vector in non-relativistic case. According to special relativity 
\begin{eqnarray}
v_x^2+v_y^2+v_z^2\leq c^2.
\label{constraint}
\end{eqnarray}
where  $v^i=\frac{d x^i}{dt}$, is the co-ordinate velocity.
In the non-relativistic case (where $v_x, v_y, v_z $ are not constraint by the above inequality), 
we write the probability of finding the particle simultaneously in the velocity range 
$v_x$ to $v_x+dv_x$, $v_y$ to $v_y+dv_y$, and, $v_z$ to $v_z+ dv_z$ as the product of the individual probabilities:
\begin{eqnarray}
p(v_x, v_y, v_z)dv_x dv_y dv_z= f(v_x)f(v_y)f(v_z)dv_xdv_ydv_z\ \ \ \ \ 
\label{cmbprbl}
\end{eqnarray} 
where $f(v_i)dv_i$ represents the probability of finding the particle in the velocity component range $v_i$ to $v_i+dv_i$.
The functional form of `$f$' is same for all the components due to isotropy. This assumptions ($\ref{cmbprbl}$)
is no longer valid in relativistic case
as the co-ordinate velocity component are no longer independent due the above inequality and can only vary from $-c$ to $c$. 
This problem can be circumvented if we seek a formula for the probability distribution of 4-velocity instead of coordinate velocity. 
The 4-velocity components are related through the equation
\begin{eqnarray}  
\sum_{i=1}^3(u^i)^2=c^2[(u^0)^2-1]=v^2/(1-\frac{v^2}{c^2}).
\end{eqnarray}
But, as $v\rightarrow c$, $u^{0}\rightarrow \infty$ and therefore, each of the three spacial components ($u^1\!, u^2\!, u^3$) 
of the 4-velocity can take values from $-\infty$ to $\infty$ independently, 
irrespective of the values of other two components. Therefore,  
just as the 3-velocity components in the non-relativistic case, 
the three spacial components of 4-velocities have no constraint among themselves. 
In the chosen Lorentz frame, the center of mass of the gas system is 
at rest. The average 4-momentum of the total gas system is zero in all directions. This implies that 
the probability distribution function of the three spacial components of the 4-velocity is isotropic just as 
the probability distribution function for velocities in non-relativistic case. 
Let the probability of finding a gas molecule in between $u^i$ to $u^i+du^i$ be
\begin{eqnarray}
p_i(u^i)du^i=F_i(u^i)du^i, \ \ \  \ (i=1,2,3).
\label{4prbl1}
\end{eqnarray} 
The assumption of isotropy demands that the functional from of $F_i$ is same for all values of $i$, {\it i.e.} 
$F_1=F_2=F_3$. As already explained above the three spatial components of 4-velocity $u^1, \ u^2, \ u^3$ are not
constrained, hence the probability that a gas 
particle is found simultaneously within the 4-velocity range $u^1$ to $u^1+du^1$, $u^2$ to $u^2+du^2$, and 
$u^3$ to $u^3+du^3$ is 
\begin{eqnarray}
p(u^1\!,u^2\!,u^3)du^1du^2du^3\!=\!F(u^1)F(u^2)F(u^3)du^1du^2du^3.\ \ \ \ \
\label{4prbl2}
\end{eqnarray}
But, on account of isotropy,  $p(u^1, u^2, u^3)$ is invariant under a three dimensional rotation and 
 must be a function of $(u^1)^2+(u^2)^2+(u^3)^2=(u)^2$. This gives 
\begin{eqnarray}
F(u^1)F(u^2)F(u^3)=\psi((u)^2).
\label{dife}
\end{eqnarray}    
Differentiating Eq.{\ref{dife}} partially, w.r.t $u^1$, 
$u^2$, and $u^3$, and dividing by \!$F(u^1)F(u^2)F(u^3)$, one arrives at the equation 
\begin{eqnarray}
\frac{F^{\prime}(u^1)}{u^1F(u^1)}\!=\!\frac{F^{\prime}(u^2)}{u^2F(u^2)}\!=\!\frac{F^{\prime}(u^3)}{u^3F(u^3)}=\frac{\psi^{\prime}(u)}{u\psi}=-2 \lambda
\label{feqn}
\end{eqnarray}
where $\lambda$ is a constant which is positive because of the convergence requirement of $F$. 
The Eq.($\ref{feqn}$) is solved to give:\ 
$F(u^i)=Ae^{-\lambda {(u^i)}^2} $\ and 
therefore,
\begin{eqnarray}
F(u^1)F(u^2)F(u^3)={A}^3e^{-\lambda (u)^2}.
\end{eqnarray}
Integrating this probability over spherical shell of radius $u$ and $u+du$, one finds the total probability of finding 
the particle between the speed range $u$ and $u+du$, as 
\begin{eqnarray}
F(u)du=4\pi A^3 u^2 e^{-\lambda u^2}du.
\label{mxldstrbn}
\end{eqnarray}
The normalization condition: $\int_{0}^{\infty} F(u)du=1$ relates ${ A}$ with $\lambda$ as 
${ A}=\sqrt{\frac{\lambda}{\pi}}$. The constant $\lambda$ is related to the average energy of the gas particle 
as 
\begin{eqnarray}
\hskip -0.3cm
<\!u^0\!>=\!\!\int_{0}^{\infty}\!\!\!\!u^0F(u)du
&=&4\pi{A}^3\!\int_{0}^{\infty}\!\!\!\!u^2\sqrt{(1+\frac{u^2}{c^2})}\ e^{-\lambda u^2}du.
\nonumber
\end{eqnarray}
As in the case of MB, the parameter $\lambda$ in this case, is also related to the average kinetic energy ($<\!E_k\!>$) of the gas molecule, however not in a simple manner. If $m$ is the rest mass of the gas molecule then
\begin{eqnarray}
{<\!E_k(\lambda)\!>=mc^2[<\!u^0\!>-1]\ \ \ \ \ \ \ \ \ \ \ \ \ \ \ \ \ \ \ \ \ \ \ \ \ \ \ \ \ \ \ \ \ \ \
\nonumber}\\
=\Bigl[4\pi mc^2{ A}^3\int_{0}^{\infty}u^2\sqrt{(1+\frac{u^2}{c^2})}\ e^{-\lambda u^2}du\Bigr]-mc^2.\ \ \ \ \ \ \ 
\end{eqnarray}  
The above equation gives the physical meaning of the constant $\lambda$ as it relates this constant to the physically measurable 
quantity $<\!E_k\!>$, i.e. \!$\lambda$ is a measure of the temperature $T$ of an ideal gas.
$<\!E_k\!>$, and $T$ are monotonically decreasing function of $\lambda$. Therefore, the non-relativistic limit
is achieved in large $\lambda$ limit.
The function $F(u)$ has non-negligible value only in the region where\ $\lambda u^2 \sim 1$\ or, $u^2\sim \frac{1}{\lambda}$.
For large $\lambda$,\ this implies $u^2={v^2}/({1-\frac{v^2}{c^2}}) \ll 1$,\ i.e.\ $\frac{v^2}{c^2} \ll 1$. This gives
 $u^i \approx v^i$, therefore, $F(u)du=4\pi A^3 v^2 e^{-\lambda v^2}dv$, is the velocity distribution function in the 
non-relativistic case.

\section{ The relativistic EOS } 
In the general case of a relativistic gas, one can relate $p$, $\rho$, $\rho_0$ and 
the adiabatic index $\gamma\ (=\frac{c_p}{c_v})$ as 
\begin{eqnarray}
p=(\gamma -1)(\rho -\rho_0).
\label{gamaeqn}
\end{eqnarray} 
It is easy to show that for a cool non-relativistic gas $\gamma$ becomes $\frac{5}{3}$, and for a extreme relativistic gas of 
photon $\gamma=\frac{2}{3}$ [\onlinecite{Weinberg1972}]. 
Therefore, the value of $\gamma$ in general depends on the relativistic nature of the gas, 
{\it i.e.} the dominance of the kinetic energy density over the rest mass density $\rho_0$.    
To calculate the dependence of $\gamma$ on $\rho$, we first expressed $p$, and $\rho$, in terms of $\rho_0$. 
This can be done by considering that the pressure $p$ is the momentum flux through unit area averaged over all the molecules of 
the fluid-blob in the co-moving frame. We take a surface element of the fluid blob oriented to any one axis ($i^{th}$ axis say,) 
the pressure $p$ is\ \ 
$p=n<\!p^iv^i\!>$.
Here $n$ is the number density of the molecule,  $p^i$ is the $i^{th}$ component of the 4-momentum $p^{\mu}=mu^{\mu}$.
Using the relation $v^i=\frac{u^i}{u^0}$, and taking $<\!\!\frac{(u^1)^2}{u^0}\!\!>=<\!\!\frac{(u^2)^2}{u^0}\!\!>=<\!\!\frac{(u^3)^2}{u^0}\!\!>=<\!\!\frac{1}{3}\frac{(u)^2}{u^0}\!\!>$\ due to the isotropy, we write the pressure as 
\begin{eqnarray}
p=\frac{1}{3}\rho_0 (<\!u^0\!>- <\!\frac{1}{u^0}\!> )=\rho_0 g(\lambda).
\label{pe}
\end{eqnarray} 
Here and henceforth we work with $c=1$.
The total energy density  $\rho$ is related to rest mass density $\rho_0$ as 
\begin{eqnarray}
\rho=nm<\!u^{0}(\!\lambda\!)\!>=\rho_0 <\!u^0(\lambda)\!>=\rho_0 f(\lambda),
\label{rho}
\end{eqnarray} 
where  $f(\lambda)$ and $g(\lambda)$ are related to the modified Bessel functions of second kind $K_n(\lambda)$ through the
{\vskip -0.5cm
\begin{eqnarray}
{<\!\!u^{0}\!(\!\lambda\!)\!\!>\!=\!4\pi A^3\!\!\!\int_0^{\infty}\!\!\!\!\!\!\sqrt{1+u^2} u^2\!e^{-\lambda u^2}\!du\!=\!\frac{\pi A^3e^{\frac{\lambda}{2}}}{\lambda}K_1(\!\frac{\!\lambda\!}{2}\!),\ \ \ \ \ }
\label{au0}
\end{eqnarray}
}{\vskip -0.6cm
\begin{eqnarray}
{<\!\!\frac{1}{u^0(\!\lambda\!)}\!\!>=\!4\pi\!A^3\!\!\!\int_0^{\infty}\!\!\!\!\frac{\ e^{-\lambda u^2}\!u^2\!du}{\sqrt{1+u^2}}\!=\!\!\pi A^3\!e^{\frac{\lambda}{2}}\![K_1(\!\frac{\!\lambda\!}{2}\!)\!-\!K_0(\!\frac{\!\lambda\!}{2}\!)\!].\ \ \ \ \ }
\label{Ou0} 
\end{eqnarray}
}
The properties of $K_n(\lambda)$ functions and its derivatives (require later) are well known in mathematics ([\onlinecite{Abramowitz1972}]). The value of $4\pi A^3$ is given before using the normalization condition.

\subsection{The Specific Heat Ratio and the Sound Speed}
\begin{figure}
\includegraphics[height=6.0cm,width=8.0cm]{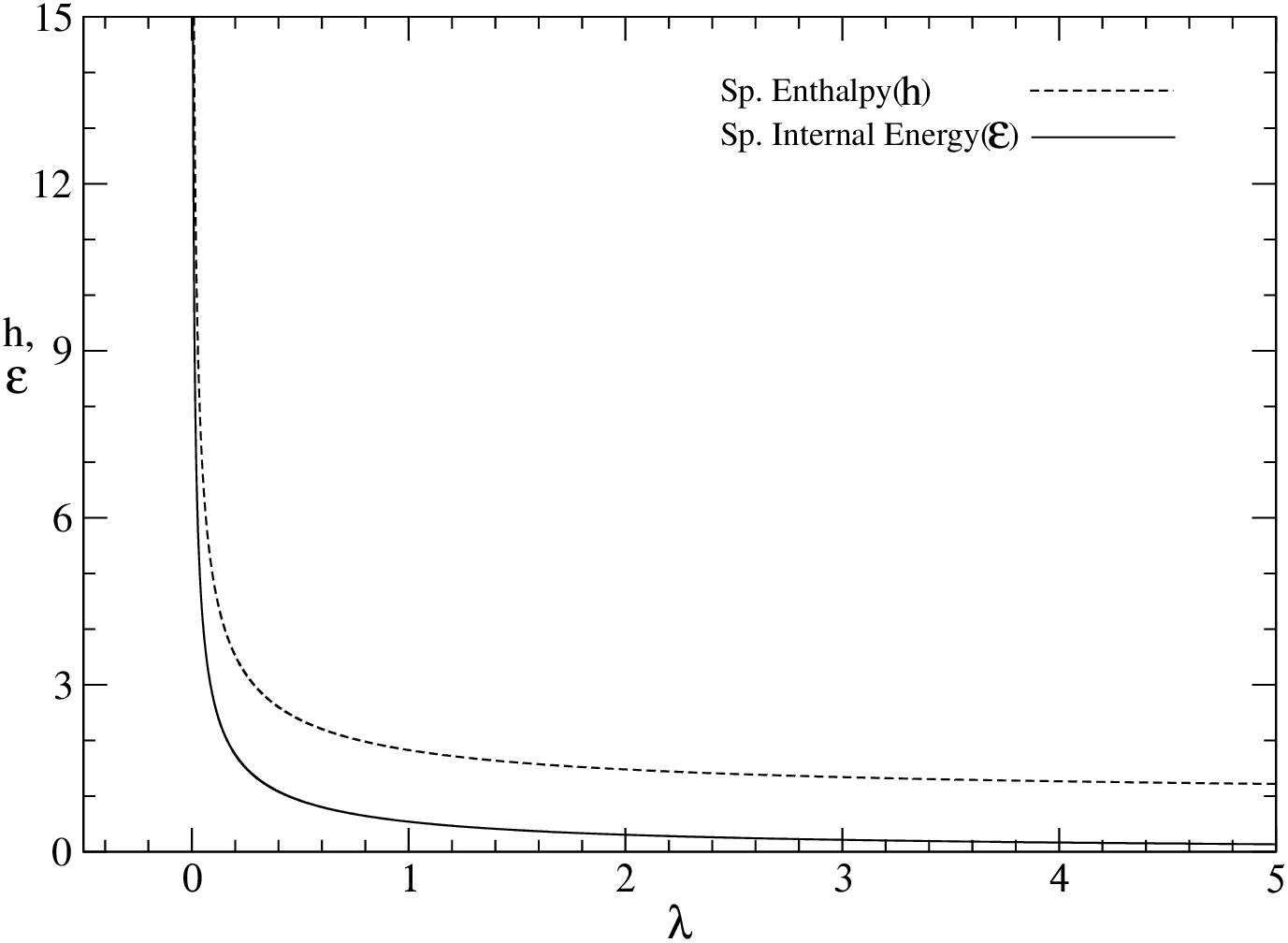}
\caption{\label{IEeps} The variation of specific internal energy ($\varepsilon$) and specific enthalpy ($h$) 
are shown as a function of $\lambda$.} 
\end{figure}

Using the EOS (\ref{pe}, \ref{rho}), one could to find the variation of $\gamma$ w.r.t $\lambda$ for a 
relativistic gas in many astrophysical applications. 
Therefore, our next job is to express the thermodynamic variables in terms of $\lambda$ alone. Using thermodynamic
definitions, we obtain the specific internal energy 
$\varepsilon=\frac{\rho-\rho_0}{\rho_0}=f-1$,\ and the specific enthalpy $h=\frac{p+\rho}{\rho_0}=f+g$ of the gas. 
Variation of $\varepsilon,\ h$  are plotted in Fig.$\ref{IEeps}$ as a function of $\lambda$.
Further, simultaneous solutions of three equations: $p=(\gamma-1)(\rho-\rho_0)$,\ $\rho=\rho_0 f(\lambda)$,\ and $p=\rho_0 g(\lambda)$,
provide the specific heat ratio
\begin{eqnarray}
\gamma=(f-1)^{-1}(h-1),
\label{gma}
\end{eqnarray} 
as function of $\lambda$ only. 
To find the expression for the sound speed, one needs to
start from the definition $a_s^2=(\frac{\partial p}{\partial \rho})_s$, together with the
entropy equation $Tds=0$ and $p=p(\rho,\rho_0)$. Computing the quantities  
$(\frac{\partial \rho_0}{\partial \rho})_s$,
$(\frac{\partial p}{\partial \lambda})_{\rho_0}$,\ $(\frac{\partial p}{\partial \rho_0})_{\lambda}$, 
and simplifying one finally yields 
\begin{eqnarray}
a_s^2\!=\!(\frac{\partial p}{\partial \rho})_{\rho_0}\!\!\left[1\!-\!(\frac{\partial \rho_0}{\partial \rho})_{\lambda}\!\right]\!+\!\frac{1}{h}\!(\frac{\partial p}{\partial \rho_0})_{\lambda}\!=\!(1\!+\!\frac{g^{\prime}}{f^{\prime}}\!)(\!\frac{g}{f\!+\!g}\!),\ \ \ \ \ \
\label{as2}
\end{eqnarray} 
in terms of $\lambda$ completely. Here the {\it prime} (`$\prime$') denotes derivatives w.r.t $\lambda$. 
To check the overall variation of $\gamma$ and $a_s$ w.r.t $\lambda$, we plot these functions in Fig.$\ref{aseps}$ and in Fig.$\ref{Gmaeps}$.
In Fig.$\ref{aseps}$, we see that there is no applicable variation of $\gamma$ throughout. 
However there is a sharp decay of $\gamma$ values very close to zero.
This indicates that EOS is mostly non-relativistic in nature and becomes relativistic to ultra-relativistic at the end phase
of evolution. On the other hand, the deviation of the sound speed $a_s$ in Fig.$\ref{Gmaeps}$, 
from non-relativistic value is prominent even in the relatively higher $\lambda$ values.
\section{Extreme limits: Ultra-relativistic and non-relativistic}
\begin{figure}
\includegraphics[height=6.0cm,width=8.0cm]{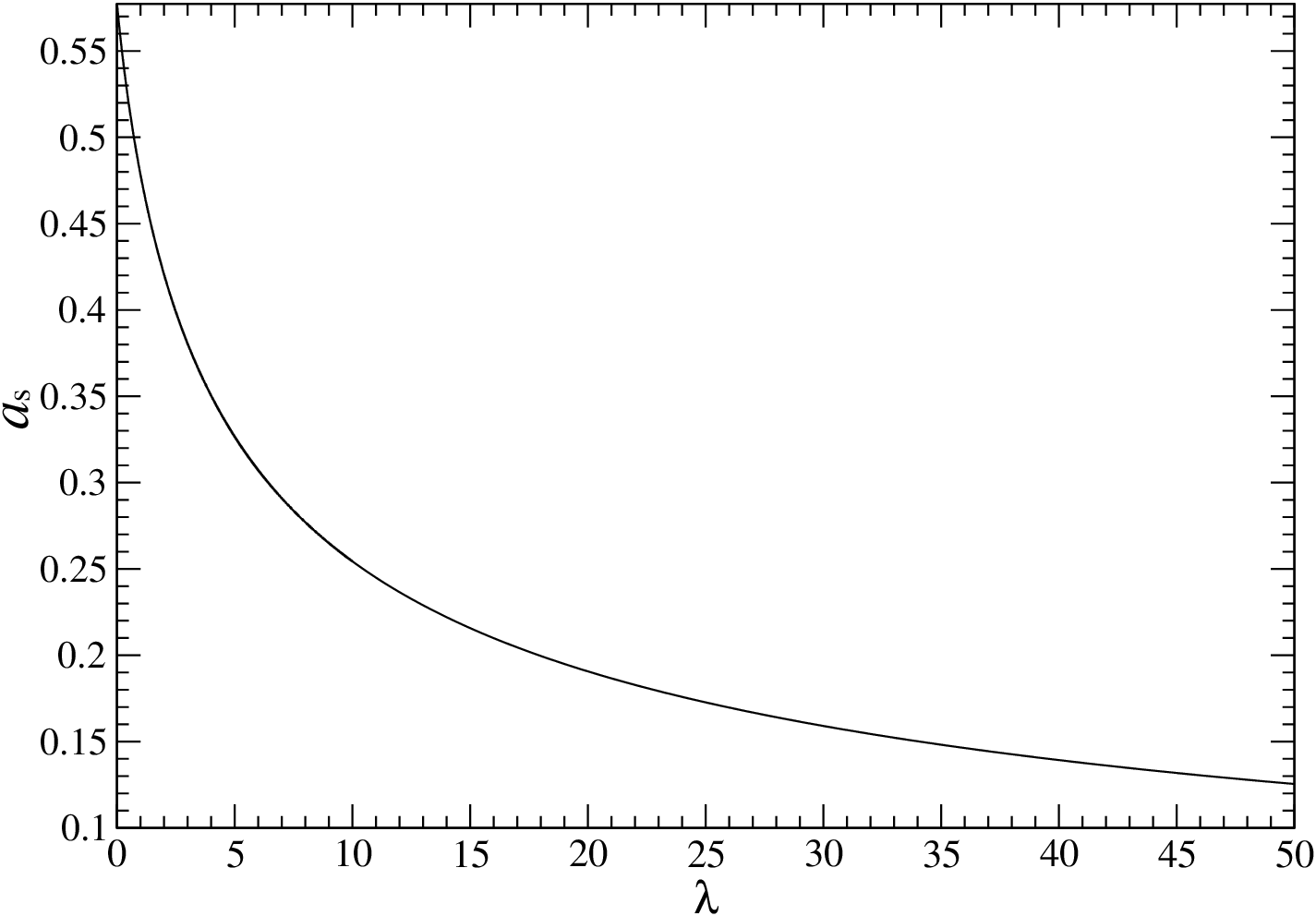}
\caption{\label{aseps} A plot of the sound speed $a_s$ vs. $\lambda$ is shown in figure.
The inequality, $a_s \le {1}/{\sqrt{3}}$ satisfies when $\lambda \to 0 $. }
\end{figure}

\begin{figure}
\includegraphics[height=6.0cm,width=8.0cm]{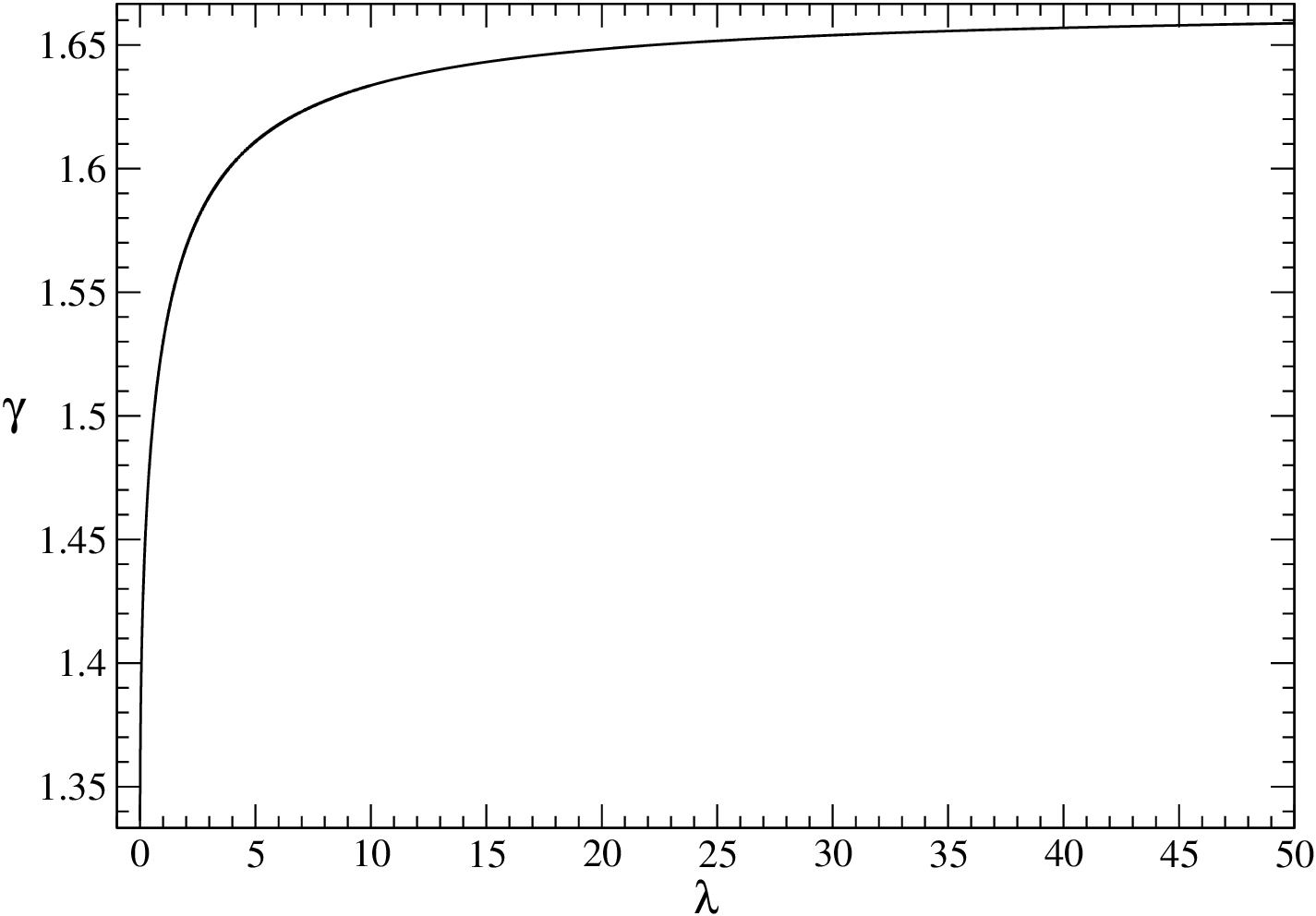}
\caption{\label{Gmaeps} The adiabatic index $\gamma$ ($\frac{4}{3}\le\gamma\le \frac{5}{3}$) is plotted w.r.t $\lambda$.} 
\end{figure}

It is important to show that this expression for $p$ gives correct results for non-relativistic 
and ultra-relativistic limit. In the {\it non-relativistic} limit $v<<c$. 
Therefore, $u^0\approx (1+\frac{1}{2}\frac{v^2}{c^2})$ and $\frac{1}{u^0}\approx (1-\frac{1}{2}\frac{v^2}{c^2})$. 
Therefore,\ 
$p=\frac{1}{3}\rho_0<\!v^2\!> $.
Now, the average kinetic energy density is\ \  $\rho_0<\!E_k\!>=\rho_0(<\!u^0\!>-1)=\frac{1}{2}\rho_0<\!v^2\!>$. Therefore, 
$\rho -\rho_0=\frac{1}{2}\rho_0<\!v^2\!>$. Using above results, we find 
\begin{eqnarray}
\rho -\rho_0=\frac{3}{2}p,
\end{eqnarray}
an exact EOS in the non-relativistic limit in which $\gamma\!=\!\frac{5}{3}$.\\
For an {\it ultra-relativistic} gas of photon 
$v \rightarrow c$ and $m\rightarrow 0$. However, $c^2mu^0=E_{photon}$ remains finite. Therefore, for a photon gas\ 
$u^0 \rightarrow \infty$ and  $\frac{1}{u^0}\rightarrow 0$. This gives\ $\rho_0=0$, $\rho=nE_{photon}$ 
and $p=\frac{1}{3}nE_{photon}$. Therefore, in the ultra-relativistic limit, EOS and $\gamma$ respectively becomes
\begin{eqnarray}
p=\frac{1}{3}\rho, \ \ \ \ \ \ \gamma=\frac{4}{3}.
\end{eqnarray}
{\vskip -0.1cm
For a gas of massive particle, the values of the $\gamma$ and $a_s$ in the extreme relativistic regime can be obtain 
from  Eq.($\ref{gma}$) and Eq.($\ref{as2}$) in the limit $\lambda \to 0$.}
The integral in Eq.($\ref{au0}$) \&  ($\ref{Ou0}$) and their derivatives w.r.t $\lambda$, 
take the value for $\lambda \to 0$ (by putting $\gamma u^2=y$)
\begin{eqnarray}
<\!\!u^0\!\!>=\!\frac{2\lambda^{-\frac{1}{2}}}{\sqrt{\pi}}, <\!\!u^0\!\!>^{\prime}=\!-\!\frac{\lambda^{-\frac{3}{2}}}{\sqrt{\pi}}, <\!\!\frac{1}{u^0}\!\!>=\!\frac{2\lambda^{\frac{1}{2}}}{\sqrt{\pi}}, <\!\!\frac{1}{u^0}\!\!>^{\prime}=\!\frac{\lambda^{-\frac{1}{2}}}{\sqrt{\pi}}.\nonumber
\end{eqnarray}
Substituting above values in the Eq.(\ref{gma}), \& (\ref{as2}), we find 
\begin{eqnarray}
\lim_{\lambda \rightarrow 0}\gamma=\frac{4}{3} \ \ \ \mbox{and} \ \ \lim_{\lambda \rightarrow 0}a_s^2=\frac{1}{3}.
\end{eqnarray}
{\vskip -0.1cm
Similarly, it can be shown that in the large $\lambda$ limit, $\gamma=\frac{5}{3}$.}
We emphasis these points in Fig.$\ref{aseps}$, where y-axis are set at $y=\frac{4}{3}$ and $\frac{5}{3}$.  
We see, $\frac{4}{3} \le \gamma \le \frac{5}{3}$ perfectly holds in our case. Also,
in the Fig.$\ref{Gmaeps}$, we find the sound speed $a_s \le \frac{1}{\sqrt{3}}$ satisfies.

\section{Conclusion}
In this communication we express the relativistic EOS in a new form through two parametric equations: 
$\rho=\rho_0 f(\lambda)$,\ and $p=\rho_0 g(\lambda)$, where $\lambda$\ is a parameter related to kinetic energy of the gas.
The EOS is obtained by using a 4-velocity distribution function which we derived by applying the basic principles of probability 
theory with the assumptions of isotropy.
In the ultra-relativistic regime,
these new equations perfectly produces well-known results, namely, the EOS:\ $\rho=3p$,\ 
the value of $\gamma=\frac{4}{3}$, and sound speed $\le \frac{1}{\sqrt{3}}$ have\\
\\
an exact match,whereas, in the non-relativistic regime, EOS correctly reduces to $\rho-\rho_0=\frac{3}{2}p$, which
implies $\gamma = \frac{5}{3}$. 
The results we obtain are also in well agreement with the known results.
Our theoretical investigation indicates that EOS evolves abruptly at the end phase of ultra-relativistic regime.
Following the prescription of new EOS, it is easy to express the thermodynamic and hydrodynamic
variables in terms of $\lambda$ alone. In the next stage, one just need to estimate the variation of $\lambda$ w.r.t 
the radial distance as to find an interesting solutions in different physical systems in interest.
We have studied one such solutions in the accretion around black hole which we wish communicated soon.\\
\begin{acknowledgments}
Author PB acknowledges SNBNCBS PDF support. Author SM acknowledges KASI PDF support and thanks to RKMRC to grant study leave under UGC-scheme.  
\end{acknowledgments}

\end{document}